\documentclass[aps,prl,letterpaper,twocolumn,longbibliography,floatfix,nofootinbib]{revtex4-2}

\usepackage[utf8]{inputenc}
\usepackage{graphicx}
\usepackage{dcolumn}
\usepackage{bm}
\usepackage{bbm}
\usepackage{hyperref}
\usepackage{xcolor}
\usepackage{amsmath}
\usepackage{amssymb}

\begin{document}

\title{A triadic approximation reveals the role of interaction overlap on the spread of complex contagions on higher-order networks}

\author{Giulio Burgio}
\author{Sergio G\'omez}
\author{Alex Arenas}

\affiliation{
Departament d'Enginyeria Inform\`atica i Matem\`atiques, Universitat Rovira i Virgili, 43007 Tarragona, Spain
}


\begin{abstract}
Contagion processes relying on the exposure to multiple sources are prevalent in social systems, and are effectively represented by hypergraphs. In this Letter, we derive a mean-field model that goes beyond node- and pair-based approximations. We reveal how the stability of the contagion-free state is decided by either two- or three-body interactions, and how this is strictly related to the degree of overlap between these interactions. Our findings demonstrate the dual effect of increased overlap: it lowers the invasion threshold, yet produces smaller outbreaks. Corroborated by numerical simulations, our results emphasize the significance of the chosen representation in describing a higher-order process.
\end{abstract}

\maketitle

\paragraph{Introduction.}

Behaviors, strategies, and conventions often require some form of reinforcement for their adoption~\cite{granovetter1978threshold,watts2002simple,dodds2004universal,centola2007complex,Guilbeault2018,romanczuk2023phase}. Spreading among individuals through learning and imitation, their diffusion can be studied as a contagion process~\cite{goffman1964generalization,daley1964epidemics,levy1993contagion,hill2010infectious}. Differently from epidemics, however, repeated exposures to the same contagious source (e.g., an individual with a given behavior) are not always sufficient for transmission and multiple sources are required, defining a complex contagion~{\cite{granovetter1978threshold,centola2007complex,centola2010spread,o2015mathematical,Guilbeault2018,guilbeault2021topological}}. When transmission depends on the exposures being simultaneous{, as when interacting in a group,} a spreading event becomes a many-body interaction~\cite{iacopini2019simplicial}. Similarly, in biochemical systems, oftentimes a species needs simultaneous exposure to one or more other species in order for the reaction to occur~\cite{klamt2009hypergraphs,jost2019hypergraph}.

Here we focus on social contagion~\cite{tarde1890laws,blumer1939,le2002crowd,granovetter1978threshold,dodds2004universal,hill2010infectious}. This can be mapped to a susceptible-infectious-susceptible (SIS) process on a hypergraph~\cite{iacopini2019simplicial,landry2020effect,burgio2021network}, {a generalization of a graph where an $n$-edge (i.e., an edge incident on $n$ nodes) is used to represent an $n$-body interaction~\cite{battiston2020networks}}. We assume that a susceptible node (individual) becomes infected at rate $\beta^{(1)}$ in a two-body interaction ($2$-edge) with an infected individual ($S+I \xrightarrow{\beta^{(1)}} 2I$), at rate $\beta^{(2)}$ in a three-body interaction ($3$-edge) with two infected individuals ($S+2I \xrightarrow{\beta^{(2)}} 3I$), and so on for larger groups. Infected individuals recover at rate $\mu$ ($I \xrightarrow{\mu} S$). Within this higher-order setting, Iacopini \emph{et al.}~\cite{iacopini2019simplicial} found a phenomenology (the appearance of a saddle-node bifurcation implying critical-mass behavior) that is effectively equivalent to the one already uncovered by Dodds \emph{et al.}~\cite{dodds2004universal}. The latter used a threshold model (where transmission generally requires multiple exposures) over two-body interactions only. One might thus believe that accounting for higher-order, group interactions is only a marginal refinement, if not an unnecessary complication.

In this Letter, we challenge this belief by demonstrating that the outcome of the contagion process is fundamentally linked to how interactions of different orders are arranged in the system. Having developed a clique-based mean-field model that accounts for local dynamical correlations, we reveal that, contrary to the predictions made by node-based approximations~\cite{iacopini2019simplicial,landry2020effect}, the invasion threshold, $\beta^{(1)}_\text{cr}$, at which the inactive (contagion-free) state becomes unstable, does depend on $\beta^{(2)}$~\cite{burgio2021network}. This dependence is proven to be strictly related to the degree of overlap between three- and two-body interactions. Having derived an explicit expression for the critical surface, we demonstrate that the overlap has a double-edged effect: it lowers the invasion threshold, but also makes the outbreaks generally smaller.

\paragraph{Triadic approximation.}

Let us start from the closure approximation we apply to the exact microscopic equations on hypergraphs. We track the state evolution of subsets of nodes which form maximal cliques (i.e., {cliques not subsets} of larger ones) in the projection graph constructed by associating cliques to edges of the hypergraph. Accordingly, considering up to three-body interactions, we account for the evolution of the state probability $P^{\sigma_i}_i$ for node $i$ to be in state $\sigma_i$, $P^{\sigma_i\sigma_j}_{ij}$ for the maximal link ${ij}$ to be in state ${\sigma_i\sigma_j}$, $P^{\sigma_i\sigma_j\sigma_l}_{ijl}$ for the (maximal) $3$-clique ${ijl}$ to be in state ${\sigma_i\sigma_j\sigma_l}$. Notice that a $3$-clique, when projected back to the hypergraph, comes in one of three flavors: a length-$3$ cycle (or $3$-cycle), conveying three two-body interactions, a $3$-edge, conveying a three-body interaction, or a $2$-simplex (or triangle), conveying all of them.

The state probability of other local structures is approximated in terms of the maximal cliques composing it. We consider random hypergraphs that are sparse to the extent that the probability for two maximal cliques to share more than one node vanishes in the infinite-size limit~\footnote{The number of maximal edges of any cardinality must scale slower than quadratically with the system size.}. We thus need a closure only for the following local structures: two connected maximal links, a maximal link connected to a $3$-clique, and two connected $3$-cliques. We approximate their state as follows~\cite{burgio2021network}
\begin{subequations}
\label{eq:MCA}
\begin{align}
    P^{\sigma_i\sigma_j\sigma_l}_{i\underline{j}l} \approx&\ P^{\sigma_i\sigma_j}_{ij}P^{\sigma_j\sigma_l}_{jl}/P^{\sigma_j}_j\ , \label{eq:MCA1}\\
    P^{\sigma_i\sigma_j\sigma_l\sigma_h}_{i\underline{j}lh} \approx&\ P^{\sigma_i\sigma_j}_{ij}P^{\sigma_j\sigma_l\sigma_h}_{jlh}/P^{\sigma_j}_j\ , \label{eq:MCA2}\\
    P^{\sigma_i\sigma_j\sigma_l\sigma_h\sigma_k}_{ij\underline{l}hk} \approx&\ P^{\sigma_i\sigma_j\sigma_l}_{ijl}P^{\sigma_l\sigma_h\sigma_k}_{lhk}/P^{\sigma_l}_l\ , \label{eq:MCA3}
\end{align}
\end{subequations}
where the underline indicates the shared node. We refer to Eqs.~(\ref{eq:MCA}) as the \emph{triadic approximation}~\footnote{Notice that this closure becomes exact for non-recursive dynamics such as susceptible-infected-recovered (SIR) processes when the structure has no cycles of more than three nodes~\cite{kiss2017mathematics}.}.

The higher-order interaction structure is encoded in the following binary tensors: $A^{(1)}$, such that $A^{(1)}_{ij} = 1$ if the maximal link $ij$ exists; $A^{(1,0)}$ and $A^{(0,1)}$, such that $A^{(1,0)}_{ijl} = 1$ ($A^{(1,0)}_{ijl} = 0$) and $A^{(0,1)}_{ijl} = 0$ ($A^{(0,1)}_{ijl} = 1$) if $ijl$ is a $3$-cycle ($3$-edge); and such that $A^{(1,0)}_{ijl}A^{(0,1)}_{ijl} = 1$ if $ijl$ is a triangle (for later convenience, we introduce $A^{(1,1)}=A^{(1,0)}\odot A^{(0,1)}$). Specifically, if for any $3$-clique $ijl$, $A^{(0,1)}_{ijl} = 1 \Rightarrow A^{(1,0)}_{ijl} = 1$, the hypergraph is a simplicial $2$-complex{, for the existence of a $3$-edge implies the existence of the $2$-edges it includes}~\cite{bretto2013hypergraph}. If, instead, $A^{(0,1)}_{ijl} = 1 \Rightarrow A^{(1,0)}_{ijl} = 0$, it is a linear hypergraph, for any two edges will share at most one node~\cite{bretto2013hypergraph}. Any hypergraph is located in between these two limits, depending on the degree of overlap between three- and two-body interactions.

Having rescaled time by $\mu$, the process is described by the following system of microscopic equations,
\begin{widetext}
\begin{subequations}
\label{eq:micro}
\begin{align}
    \dot{P}^I_i =& -P^I_i + \beta^{(1)} \sum_j A^{(1)}_{ij}P^{SI}_{ij} + \frac12\sum_{j,l}  \left[A^{(1,0)}_{ijl}\beta^{(1)}(P^{SSI}_{ijl}+P^{SIS}_{ijl}+2P^{SII}_{ijl})+A^{(0,1)}_{ijl}\beta^{(2)}P^{SII}_{ijl}\right]\ ,
    \\
    \dot{P}^{SI}_{ij} =& -(1+\beta^{(1)})P^{SI}_{ij} + P^{II}_{ij} - \beta^{(1)} \sum_{l\neq j} A^{(1)}_{il}P^{ISI}_{j\underline{i}l} + \beta^{(1)} \sum_{l\neq i} A^{(1)}_{jl}P^{SSI}_{i\underline{j}l} \notag \\
    &- \frac12\sum_{l,h} \left[A^{(1,0)}_{ilh}\beta^{(1)}(P^{ISIS}_{j\underline{i}lh}+P^{ISSI}_{j\underline{i}lh}+2P^{ISII}_{j\underline{i}lh})+A^{(0,1)}_{ilh}\beta^{(2)}P^{ISII}_{j\underline{i}lh}\right] + \{i\leftrightarrow j\}\ ,
    \\
    \dot{P}^{SSI}_{ijl} =& -(1+2A^{(1,0)}_{ijl}\beta^{(1)})P^{SSI}_{ijl} + P^{ISI}_{ijl} + P^{SII}_{ijl} \notag \\
    &- \beta^{(1)} \sum_{h\neq j,l} A^{(1)}_{ih}P^{SISI}_{jl\underline{i}h} - \beta^{(1)} \sum_{h\neq i,l} A^{(1)}_{jh}P^{SISI}_{il\underline{j}h} + \beta^{(1)} \sum_{h\neq i,j} A^{(1)}_{lh}P^{SSSI}_{ij\underline{l}h} \notag \\
    &- \frac12\sum_{h,k\neq j,l} \left[A^{(1,0)}_{ihk}\beta^{(1)}(P^{SISIS}_{jl\underline{i}hk}+P^{SISSI}_{jl\underline{i}hk}+2P^{SISII}_{jl\underline{i}hk})+A^{(0,1)}_{ihk}\beta^{(2)}P^{SISII}_{jl\underline{i}hk}\right] - \{i\leftrightarrow j\} + \{i\leftrightarrow l\}\ ,
    \\
    \dot{P}^{SII}_{ijl} =& -(2+2A^{(1,0)}_{ijl}\beta^{(1)}+A^{(0,1)}_{ijl}\beta^{(2)})P^{SII}_{ijl} + A^{(1,0)}_{ijl}\beta^{(1)}(P^{SSI}_{ijl} + P^{SIS}_{ijl}) + P^{III}_{ijl} \notag \\
    &- \beta^{(1)} \sum_{h\neq j,l} A^{(1)}_{ih}P^{IISI}_{jl\underline{i}h} + \beta^{(1)} \sum_{h\neq i,l} A^{(1)}_{jh}P^{SISI}_{il\underline{j}h} + \beta^{(1)} \sum_{h\neq i,j} A^{(1)}_{lh}P^{SISI}_{ij\underline{l}h} \notag \\
    &- \frac12\sum_{h,k\neq j,l} \left[A^{(1,0)}_{ihk}\beta^{(1)}(P^{IISIS}_{jl\underline{i}hk}+P^{IISSI}_{jl\underline{i}hk}+2P^{IISII}_{jl\underline{i}hk})+A^{(0,1)}_{ihk}\beta^{(2)}P^{IISII}_{jl\underline{i}hk}\right] +\{i\leftrightarrow j\} + \{i\leftrightarrow l\}\ ,
\end{align}
\end{subequations}
\end{widetext}
\noindent where $\{i\leftrightarrow j\}$ denotes that obtained by swapping $i$ and $j$ in the explicit term (excluding the sign in front) on the same line and taking $i$ and $j$ in state $S$. The other state probabilities are found as $P^S_i = 1 - P^I_i$, $P^{SS}_{ij} = 1 - P^I_i - P^{SI}_{ij}$, $P^{II}_{ij} = P^I_{i} - P^{IS}_{ij}$, $P^{SSS}_{ijl} = 1 - P^I_i - P^{SII}_{ijl} - P^{SSI}_{ijl} - P^{SIS}_{ijl}$, $P^{III}_{ijl} = P^I_i - P^{ISI}_{ijl} - P^{IIS}_{ijl} - P^{ISS}_{ijl}$. Equations~(\ref{eq:micro}) are closed through Eqs.~(\ref{eq:MCA}). The system consists then of $N+L+2(T^{(1,0)}+T^{(0,1)}+T^{(1,1)})$ equations, being $N$, $L$, $T^{(1,0)}$, $T^{(0,1)}$, and $T^{(1,1)}$, the number of nodes, maximal links, $3$-cycles, $3$-edges, and triangles, respectively.

\paragraph{Mean-field approximation.}

\begin{figure}
    \centering
    \includegraphics[width = 1.\linewidth]{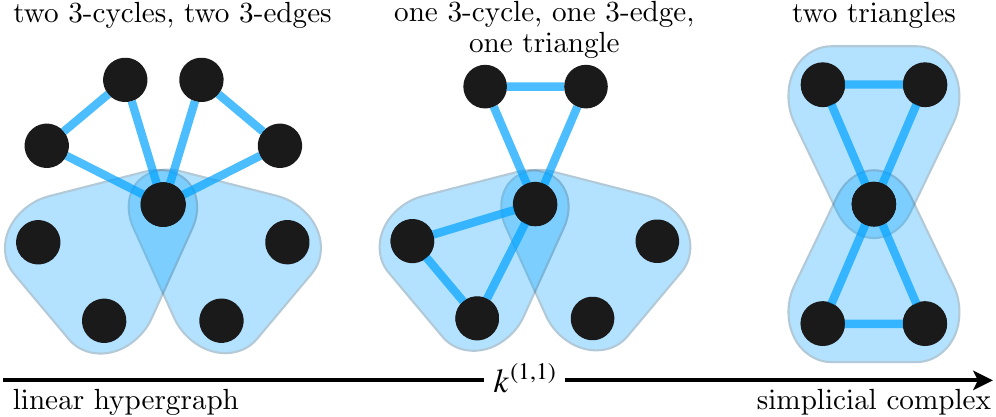}
    \caption{Example showing how the neighborhood of a focal node changes with $k^{(1,1)}$ for fixed $\kappa^{(1)}$, $k^{(1)}$ (zero here) and $\kappa^{(2)}$. The node takes part to $\kappa^{(1)}=4$ two-body
    and $\kappa^{(2)}=2$ three-body interactions, but their degree of overlap changes.
    }
    \label{fig_1}
\end{figure}

To make this model analytically tractable, we perform a mean-field approximation by regarding all the nodes and cliques as equivalent to their average counterparts. Accordingly, every node is assumed to be part of the same number of maximal links $k^{(1)}$---$3$-cliques, $k^{(1,0)}$; $3$-edges, $k^{(0,1)}$; and triangles, $k^{(1,1)}$---and thus participates in $\kappa^{(1)}=k^{(1)}+2(k^{(1,0)}+k^{(1,1)})$ two-body interactions and {$\kappa^{(2)}=k^{(0,1)}+k^{(1,1)}$} three-body interactions [see Fig.~\ref{fig_1}]. The state probabilities $P^\sigma_i$, $P^{\sigma\sigma^\prime}_{ij}$ and $P^{\sigma\sigma^\prime\sigma^{\prime\prime}}_{ijl}$,  with $\sigma,\sigma^\prime,\sigma^{\prime\prime}\in\{S,I\}$, are taken equal to their respective averages, $P^\sigma = \sum_i P^\sigma_i/N$, $P^{\sigma\sigma^\prime} = \sum_{i,j} A^{(1)}_{ij}P^{\sigma\sigma^\prime}_{ij}/Nk^{(1)}$, and $P^{\sigma\sigma^\prime\sigma^{\prime\prime}}_{x} = \sum_{i,j,l} A^{x}_{ijl}P^{\sigma\sigma^\prime\sigma^{\prime\prime}}_{ijl}/2Nk^{x}$, the index $x\in\{(1,0),(0,1),(1,1)\}$ indicating the type of the considered $3$-clique~\footnote{Notice that permuting the superscripts $\sigma$, $\sigma^\prime$, and $\sigma^{\prime\prime}$, has no effect.}.

Using the indicator function ${\mathbbm{1}}_p$, giving $1$ if condition $p$ is fulfilled and $0$ otherwise, the reduced system reads
\begin{widetext}
\begin{subequations}
\label{eq:MF}
\begin{align}
    \dot{P}^I =& -P^I + \beta^{(1)} k^{(1)} P^{SI} + 2\beta^{(1)} \left[k^{(1,0)}(P^{SSI}_{(1,0)}+P^{SII}_{(1,0)}) + k^{(1,1)}(P^{SSI}_{(1,1)}+P^{SII}_{(1,1)})\right] + \beta^{(2)} \left[k^{(0,1)}P^{SII}_{(0,1)}+k^{(1,1)}P^{SII}_{(1,1)}\right]\ ,
    \label{eq:MF1} \\
    \dot{P}^{SI} =& -(1+\beta^{(1)}) P^{SI} + P^{II} - \beta^{(1)} (k^{(1)}-1) P^{SI}\frac{P^{SI}-P^{SS}}{P^S} \notag \\
    &- \left\{2\beta^{(1)} \left[k^{(1,0)}(P^{SSI}_{(1,0)}+P^{SII}_{(1,0)}) + k^{(1,1)}(P^{SSI}_{(1,1)}+P^{SII}_{(1,1)})\right] + \beta^{(2)} \left[k^{(0,1)}P^{SII}_{(0,1)}+k^{(1,1)}P^{SII}_{(1,1)}\right]\right\}\frac{P^{SI}-P^{SS}}{P^S}\ ,
    \label{eq:MF2} \\
    \dot{P}^{SSI}_{x} =& -2(1+\beta^{(1)}{\mathbbm{1}}_{x\neq(0,1)}) P^{SSI}_{x} + 2P^{SII}_{x} -\beta^{(1)}k^{(1)}P^{SI}\frac{2P^{SSI}_{x}-P^{SSS}_{x}}{P^S} \notag \\
    &- 2\beta^{(1)}\left[(k^{(1,0)}-{\mathbbm{1}}_{x=(1,0)})(P^{SSI}_{(1,0)}+P^{SII}_{(1,0)}) + (k^{(1,1)}-{\mathbbm{1}}_{x=(1,1)})(P^{SSI}_{(1,1)}+P^{SII}_{(1,1)})\right]\frac{2P^{SSI}_{x}-P^{SSS}_{x}}{P^S} \notag \\
    &- \beta^{(2)}\left[(k^{(0,1)}-{\mathbbm{1}}_{x=(0,1)})P^{SII}_{(0,1)}+(k^{(1,1)}-{\mathbbm{1}}_{x=(1,1)})P^{SII}_{(1,1)}\right]\frac{2P^{SSI}_{x}-P^{SSS}_{x}}{P^S}\ ,
    \label{eq:MF3} \\
    \dot{P}^{SII}_{x} =& -(2+2\beta^{(1)}{\mathbbm{1}}_{x\neq(0,1)} + \beta^{(2)}{\mathbbm{1}}_{x\neq(1,0)}) P^{SII}_{x} +2\beta^{(1)}{\mathbbm{1}}_{x\neq(0,1)}P^{SSI}_{x} + P^{III}_{x} -\beta^{(1)}k^{(1)}P^{SI}\frac{P^{SII}_{x}-2P^{SSI}_{x}}{P^S} \notag \\
    &- 2\beta^{(1)}\left[(k^{(1,0)}-{\mathbbm{1}}_{x=(1,0)})(P^{SSI}_{(1,0)}+P^{SII}_{(1,0)}) + (k^{(1,1)}-{\mathbbm{1}}_{x=(1,1)})(P^{SSI}_{(1,1)}+P^{SII}_{(1,1)})\right]\frac{P^{SII}_{x}-2P^{SSI}_{x}}{P^S} \notag \\
    &- \beta^{(2)}\left[(k^{(0,1)}-{\mathbbm{1}}_{x=(0,1)})P^{SII}_{(0,1)}+(k^{(1,1)}-{\mathbbm{1}}_{x=(1,1)})P^{SII}_{(1,1)}\right]\frac{P^{SII}_{x}-2P^{SSI}_{x}}{P^S}\ ,
    \label{eq:MF4}
\end{align}
\end{subequations}
\end{widetext}
where $P^S = 1 - P^I$, $P^{SS} = 1 - P^I - P^{SI}$, $P^{II} = P^I - P^{SI}$, $P^{SSS} = 1 - P^I - P^{SII} - 2P^{SSI}$, $P^{III} = P^I - P^{SSI} - 2P^{SII}$.

\begin{figure*}[t]
    \centering
    \includegraphics[width = 0.85\linewidth]{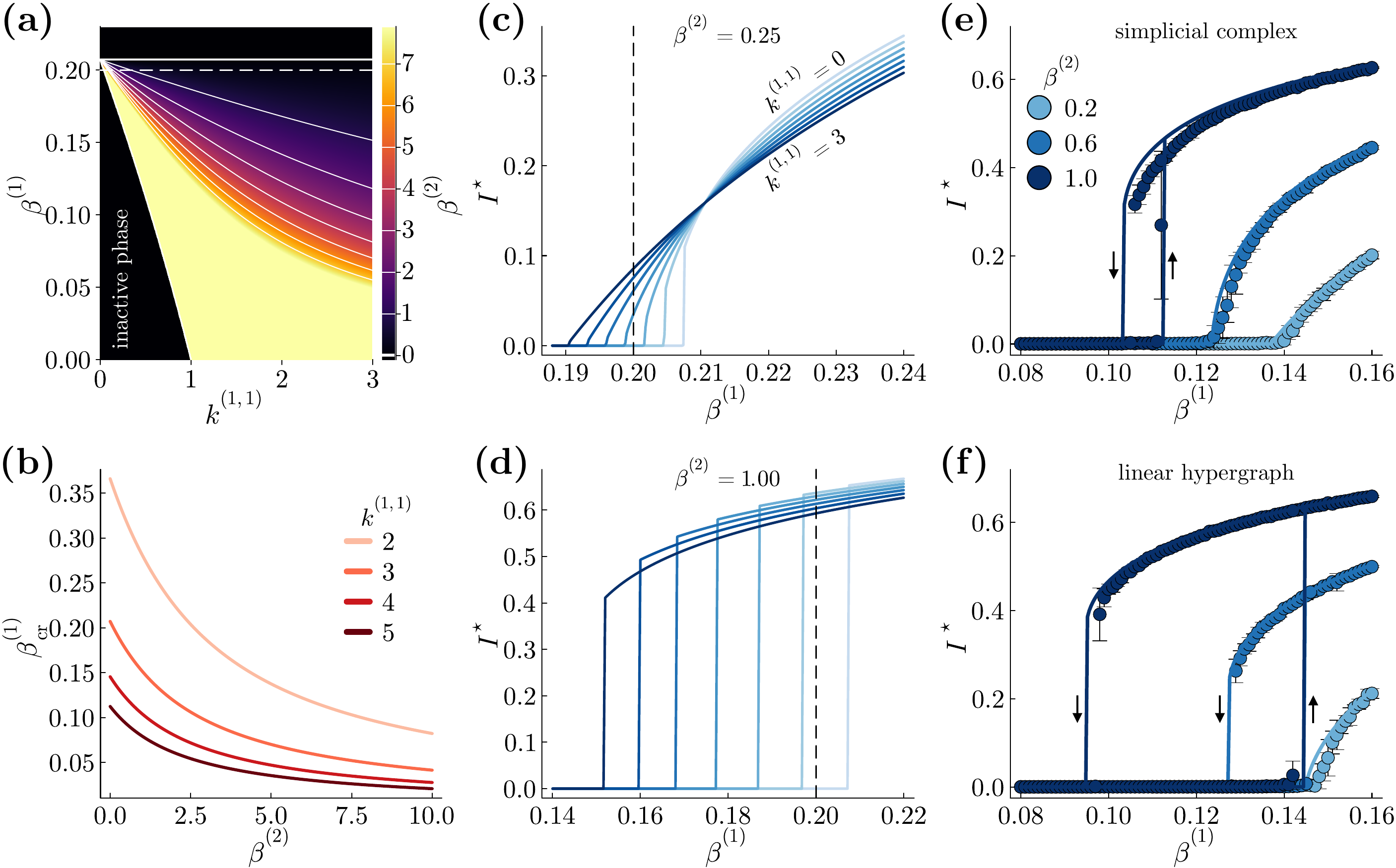}
    \caption{Predictions from the mean-field model. (a) Critical surface $(k^{(1,1)},\beta^{(1)},\beta^{(2)})_\text{cr}$ defined by Eq.~(\ref{eq:critical_cond}) for hypergraphs with $\kappa^{(1)}=6$ two-body ($k^{(1)}=0$) and $\kappa^{(2)}=3$ three-body interactions per node. The white solid curves correspond to the values of $\beta^{(2)}$ indicated in the color bar (limited at $7$ for better readability), the curve $\beta^{(2)}=0$ being thicker. The dashed line denotes $\beta^{(1)}_{(1)}=1/5$, associated to the locally treelike network with the same $\kappa^{(1)}$. (b) Critical threshold $\beta^{(1)}_\text{cr}\equiv \beta^{(1)}_{(1,1)}$, Eq.~(\ref{eq:critical_point}), for homogeneous simplicial $2$-complexes. (c)--(d) Equilibrium prevalence, $I^\star$, for $\beta^{(2)} \in \{0.25,1.00\}$, and $k^{(1,1)}$ from $0$ (lightest shade) to $3$ (darkest shade) in steps of $0.5$. The dashed line indicates $\beta^{(1)}_{(1)}=1/5$. (e)--(f) {Comparison of the model (solid lines) with numerical simulations} performed on random regular hypergraphs with $N=5000$, $\kappa^{(1)}=8$, $k^{(1)}=2$, $\kappa^{(2)}=3$, and $k^{(1,1)}=3$ (simplicial complex) and $k^{(1,1)}=0$ (linear hypergraph). Points and error bars denote averages and standard errors computed over $20$ random initializations. The arrows help to distinguish the forward and backward curves in hysteresis cycles.
    }
    \label{fig_2}
\end{figure*}

To correctly locate the phase transition, we linearize Eqs.~(\ref{eq:MF}) around the {inactive state} by regarding of the same order $\epsilon \ll 1$ the probabilities of infected states, i.e., $P^{I},P^{SI},P^{II},P^{SSI}_{x},P^{SII}_{x},P^{III}_{x}\in{\cal O}(\epsilon)$~\cite{burgio2021network}. The rightmost eigenvalue of the Jacobian matrix associated with the resulting linear system crosses the imaginary axis when the following is satisfied:\footnote{This can be more readily found by imposing stationarity in Eqs.~(\ref{eq:MF}) and solving the resulting homogeneous system.}
\begin{widetext}
\begin{equation}
\label{eq:critical_cond}
         k^{(1)} \frac{\beta^{(1)}}{1+\beta^{(1)}} + k^{(1,0)} \frac{2\beta^{(1)}(1+\beta^{(1)})}{1+2\beta^{(1)}(1+\beta^{(1)})} + k^{(1,1)} \frac{\beta^{(1)}(2+2\beta^{(1)}+\beta^{(2)})}{1+\beta^{(1)}(2+2\beta^{(1)}+\beta^{(2)})} = 1\ .
\end{equation}
\end{widetext}
Equation~(\ref{eq:critical_cond}) defines the critical surface in the parameter space [see Fig.~\ref{fig_2}(a)]. There is no solution when either $\beta^{(1)} = 0$ or there are only three-body interactions ($\kappa^{(2)} = k^{(0,1)}$, $\kappa^{(1)} = 0$). This reveals that a three-body interaction cannot affect the stability of the {inactive state} unless ``activated'' by the presence of two-body interactions within the same subset of nodes. Two-body interactions are thus needed to destabilize the {inactive state}. That activation occurs in triangles. Since each term in the l.h.s.\ of Eq.~(\ref{eq:critical_cond}) is a strictly increasing function of the infection rates, given $k^{(1,1)}>0$, the larger is {$\beta^{(2)}$ ($\beta^{(1)}$), the smaller is $\beta^{(1)}=\beta^{(1)}_\text{cr}$ ($\beta^{(2)}=\beta^{(2)}_\text{cr}$) solving Eq.~(\ref{eq:critical_cond}). In particular, when triangles can percolate the structure (i.e., $k^{(1,1)} > 1$), $\beta^{(1)}_\text{cr}$ can be made arbitrarily small by increasing $\beta^{(2)}$. Moreover, by imposing $\beta^{(2)} = 0$ in Eq.~(\ref{eq:critical_cond}), we find that a simple contagion suffices to cause extensive spreads at $\beta^{(1)}_\text{cr}\in[\beta^{(1)}_{(1)},\beta^{(1)}_{(1,0)}]$}, being $\beta^{(1)}_{(1)}=1/(\kappa^{(1)}-1)$ and $\beta^{(1)}_{(1,0)}=[\sqrt{1+4/(\kappa^{(1)}-2)}-1]/2$ the critical point for, respectively, a locally treelike network ($\kappa^{(2)} = 0$, $\kappa^{(1)} = k^{(1)}$) and a 3-cycle-based network ($\kappa^{(2)} = 0$, $\kappa^{(1)} = 2k^{(1,0)}$). In agreement with previous studies proving clustering to raise the critical point of simple ~\cite{miller2009percolation,hebert2010propagation} and slightly nonlinear~\cite{keating2022multitype} contagions, $\beta^{(1)}_{(1,0)} > \beta^{(1)}_{(1)}$.

Considering then a triangle-based network ($\kappa^{(2)} = k^{(1,1)}$, $\kappa^{(1)} = 2k^{(1,1)}$), i.e., a homogeneous simplicial $2$-complex, we find that the critical point, $\beta^{(1)}_{(1,1)}$, reads
\begin{equation}
    \beta^{(1)}_{(1,1)} = \frac{\beta^{(2)}+2}{4}\left[\sqrt{1 + \frac{16}{(\kappa^{(1)}-2)(\beta^{(2)}+2)^2}} - 1\right]\ .
    \label{eq:critical_point}
\end{equation}
$\beta^{(1)}_{(1,1)}$ thus vanishes as $1/\beta^{(2)}$ for large $\beta^{(2)}$ [see Fig.~\ref{fig_2}(b)]. Observe that Eq.~(\ref{eq:critical_point}) reflects the fact that extensive contagions are possible only for $\kappa^{(1)} > 2$ ($k^{(1,1)} > 1$), when a giant connected component can exist.

To isolate the effect of the overlap between two- and three-body interactions, we fix $\kappa^{(1)}$, $k^{(1)}$ and $\kappa^{(2)}$, and increase $k^{(1,1)}$ from $0$ to $\kappa^{(2)}$ {(correspondingly, $k^{(1,0)}$ and $k^{(0,1)}$ both decrease)}. Importantly, a larger $k^{(1,1)}$ implies a smaller and more redundant neighborhood (see Fig.~\ref{fig_1}). One may thus expect that the critical point increases with $k^{(1,1)}$. As shown in detail in Fig.~\ref{fig_2}(a), instead, either $\beta^{(1)}_\text{cr}$ and $\beta^{(2)}_\text{cr}$ decrease with $k^{(1,1)}$, taking the lowest values in a simplicial complex and the highest in a linear hypergraph {(for which, being $k^{(1,1)}=0$, $\beta^{(1)}_\text{cr}$ is unaffected by $\beta^{(2)}$)}. Since $3$-edges yield a negligible contribution {around the inactive state, exchanging them for triangles helps the spread to thrive. As Figs.~\ref{fig_2}(c) and \ref{fig_2}(d) show, this holds also for the equilibrium fraction of infected nodes, $I^\star$, when $\beta^{(1)}$ is close enough to the threshold for the simplicial complex. For larger infection rates, however, that redundancy becomes detrimental, for potentially infectious edges lead to nodes which are already infected. The largest spreads are thus found for linear hypergraphs, ensuring the least-redundant, widest neighborhoods. Finally, notice in Fig.~\ref{fig_2}(c) how solely varying the overlap can change the nature of the phase transition.}

We test the model on random regular hypergraphs, generated through a standard configuration model in which every node is assigned the same degrees $k^{(1)}$, $k^{(0,1)}$, $k^{(1,0)}$, and $k^{(1,1)}$. As reported in Figs.~\ref{fig_2}(e) and \ref{fig_2}(f), numerical simulations confirm the (quantitative) predictions made by the mean-field model.

\begin{figure}[t]
    \centering
    \includegraphics[width = 0.99\linewidth]{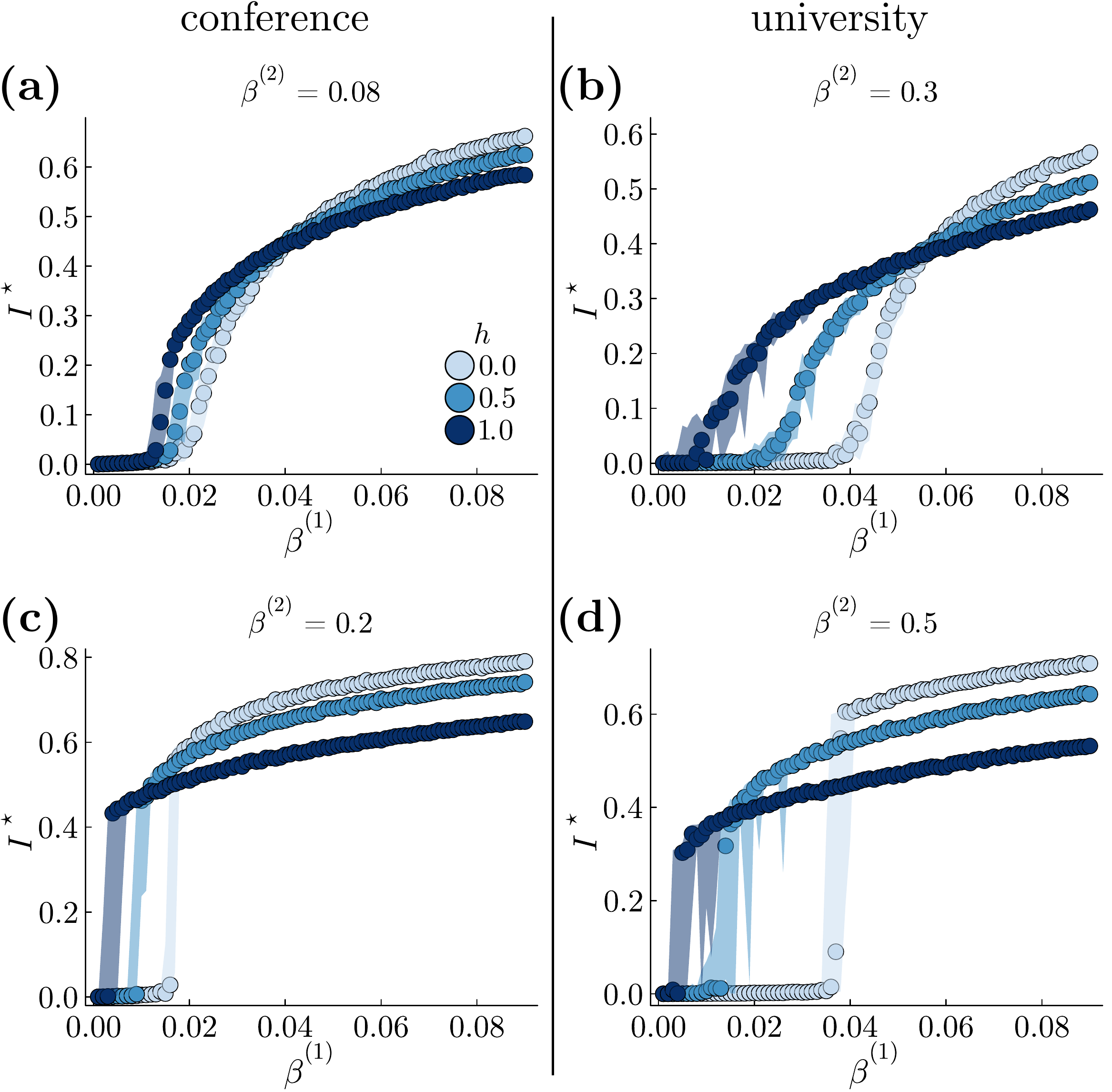}
    \caption{Numerical simulations performed on the hypergraphs constructed from the conference's~\cite{genois2018can} [(a) and (c)] and the university campus'~\cite{sapiezynski2019interaction} [(b) and (d)] datasets. Points denote medians computed over $20$ random initializations; ribbons cover from the $5$-th to the $95$-th percentile. The added three-body interactions form triangles with probability $h=0$ (linear hypergraph), $h=0.5$, and $h=1$ (simplicial complex).
    }
    \label{fig_3}
\end{figure}

We further test those predictions on hypergraphs constructed from real-world {datasets, containing record of face-to-face interactions during a conference~\cite{genois2018can}, and proximity data within a university campus~\cite{sapiezynski2019interaction}}. We refer to the Supplemental Material (SM)~\footnote{See Supplemental Material for the procedure used to generate the hypergraphs from real-world proximity data, for a detailed discussion about the quantitative limitations of the mean-field theory, and for the results found for a generalized SIR process. Refs.~\cite{d2012robustness,gleeson2012accuracy,lv2023simplicial,newman2002assortative} are included.} for the procedure used to convert each dataset into a binary network. The hypergraphs are then constructed by adding three-body interactions to either $3$-cycles with probability $h$, converting them in triangles, or to randomly selected triplets of unconnected nodes otherwise, forming $3$-edges. Even though the basic assumptions of homogeneity and sparseness that we made are heavily violated in more realistic structures (see SM), the numerical results reported in Fig.~\ref{fig_3} show that the qualitative phenomenology uncovered by the mean-field theory remains valid. This confirms its structural origin: the overlap between three- and two-body interactions. We thus conjecture that a similar picture holds for other contagion models as well, as we already verified for a SIR process (see SM).

\paragraph{Discussion.}

Through a more refined mean-field model, this study reveals a fundamental relation between the behavior of complex contagion processes and the way interactions are arranged in the higher-order structure. Extending beyond node- and pair-based approximations, our analysis establishes how three-body interactions contribute to destabilizing the inactive state, proving their contribution is contingent on overlapping with two-body interactions. Examining the boundary structures, we demonstrated that simplicial complexes and linear hypergraphs---having maximal and no overlap, respectively---exert diametrically opposed dynamical effects. The former lower the critical point, while often resulting in smaller spreads; the latter heighten the critical point, yet typically leading to larger spreads. Complementing recent findings in synchronization~\cite{zhang2023higher}, our investigation underscores the necessity of identifying the most suitable representation for specific higher-order processes.

\paragraph{Acknowledgments.}

G.B.\ acknowledges financial support from the European Union's Horizon 2020 research and innovation program under the Marie Sk\l{}odowska-Curie Grant Agreement No.\ 945413 and from the Universitat Rovira i Virgili (URV). A.A.\ and S.G.\ acknowledge support by Ministerio de Econom\'ia y Competitividad (Grants No.\ PGC2018-094754-B-C21, No.\ FIS2015-71582-C2-1, and No.\ RED2018-102518-T), Generalitat de Catalunya (Grant No.\ 2017SGR-896) and Universitat Rovira i Virgili (Grant No.\ 2019PFR-URV-B2-41). A.A.\ also acknowledges ICREA Academia, and the James S.\ McDonnell Foundation (Grant No.\ 220020325).

\clearpage

\renewcommand\theequation{{S\arabic{equation}}}
\renewcommand\thetable{{Supplementary S\Roman{table}}}
\renewcommand{\figurename}{Supplementary Figure}
\renewcommand\thefigure{{S\arabic{figure}}}
\renewcommand\thesection{{Section S\arabic{section}}}

\setcounter{section}{0}
\setcounter{table}{0}
\setcounter{figure}{0}
\setcounter{equation}{0}

\onecolumngrid

\section{Supplemental Material}

\subsection{Hypergraphs from real-world data}

\subsubsection{Generation and properties}

We first describe the procedure used to generate the hypergraphs used in Fig.~3 of the main text. We used two real-world dataset. One dataset contains face-to-face interactions recorded during a conference~\cite{genois2018can}, the other one proximity data recorded within a university campus~\cite{sapiezynski2019interaction}. The procedure is the same for both datasets.

These consist of time-resolved interactions (each representing a face-to-face interaction or proximity between two people) which, once aggregated, yields very dense networks~\cite{genois2018can}. Therefore, we first build a static pairwise network where each edge is assigned a weight equal to its number of appearances (i.e., how many times the interaction between the two agents has been detected throughout the entire observation time) and then threshold it.

Starting from an empty network with only the nodes in the dataset ($N=219$ for the conference's and $N=672$ for the university campus'), edges are listed in decreasing order of weight and added to the network starting from the first one. Since some nodes only participate to edges with very low weight, waiting until all nodes are included would yield a network identical to the original one, except for just few missing edges. To avoid this, we stop including edges when the $95\%$ of the nodes has been connected to some other node (notice that disconnected components may still exist at this point), indeed thresholding the original network\footnote{The original networks include many large cliques. Since our model assumes cliques of up to $3$ nodes, when an edge is included as above we check whether a $4$-clique formed, in which case the edge is ignored. Including $4$-cliques or larger ones does not change qualitatively the results. On the other hand, avoiding them makes the network less dense and the phenomenology easier to appreciate.}. The remaining degree-$0$ nodes are then connected to the other nodes at random. If the network is disconnected, the connected components are connected to the largest connected component by adding an edge at random between each of them and the latter. In practice, at the moment in which the $95\%$ of the nodes is reached, there exists a component containing almost all nodes and very few other components of very few nodes. Consequently, the few edges added to connect the network do not affect the properties of the thresholded network.

The binary network obtained in this way represents the backbone to which we add three-body interactions in order to get rank-$3$ hypergraphs. To do this, we first list all the $3$-cliques in the network\footnote{Notice that some $3$-cliques share two nodes with other $3$-cliques, meaning that those $2$-cliques which are part of more than one $3$-clique appear multiple times in the list. Each appearance is considered as a different interaction. An alternative method that would avoid repeated $2$-cliques consists in finding an edge-disjoint edge clique cover of the network~\cite{burgio2021network}, where, for instance, the motif made of two $3$-cliques sharing a $2$-clique would be decomposed in (and considered as) one $3$-clique (that includes the shared $2$-clique) and two $2$-cliques. Using this method just changes the number of $2$-cliques and $3$-cliques in the network, but the results remain qualitatively unaffected.}. Then, to each $3$-clique, we add a three-body interaction (i.e., a $3$-edge containing the three nodes) with probability $h$, such that, if the addition occurs, a $2$-simplex (triangle) is formed. Otherwise (occurring with probability $1-h$), the $3$-edge is added to three unconnected nodes chosen at random, so that a three-body interaction not overlapped with two-body interactions is formed. Notice that the total number of three-body interactions added is independent from $h$; only their distribution over the system changes with it. Setting $h=0$ yields a linear hypergraph. Increasing $h$, more and more frequently three-body interactions overlap with two-body interactions. At $h=1$, the structure becomes a simplicial $2$-complex.

\paragraph{Conference's dataset.}
The hypergraphs generated from this dataset consists of $N = 219$ nodes. The $2$-degree $\kappa^{(1)}$ (number of $2$-edges incident on a node) is distributed heterogeneously. The first and the second raw moments of the $2$-degree distribution are $\langle\kappa^{(1)}\rangle \approx 33.05$ and $\langle{\kappa^{(1)}}^2\rangle \approx 2034.16$, giving a high variance of $\text{var}(\kappa^{(1)}) \approx 941.84$. The structure also shows $2$-degree assortativity (coefficient $r=0.1$~\cite{newman2002assortative}). At last, the first and the second raw moments of the $3$-degree (number of $3$-edges incident on a node) distribution are $\langle\kappa^{(2)}\rangle \approx 16.22$ and $\langle{\kappa^{(2)}}^2\rangle \approx 280.03$, giving a low variance of $\text{var}(\kappa^{(2)}) \approx 16.97$.

\paragraph{University campus's dataset.}
The hypergraphs generated from this dataset consists of $N = 672$ nodes. The $2$-degree $\kappa^{(1)}$ is distributed heterogeneously. The first and the second raw moments of the $2$-degree distribution are $\langle\kappa^{(1)}\rangle \approx 15.3$ and $\langle{\kappa^{(1)}}^2\rangle \approx 479.67$, giving a high variance of $\text{var}(\kappa^{(1)}) \approx 245.56$. The structure also shows $2$-degree assortativity (coefficient $r=0.19$~\cite{newman2002assortative}). At last, the first and the second raw moments of the $3$-degree distribution are $\langle\kappa^{(2)}\rangle \approx 7.06$ and $\langle{\kappa^{(2)}}^2\rangle \approx 57.68$, giving a low variance of $\text{var}(\kappa^{(2)}) \approx 7.84$.

\subsubsection{Comparison with the mean-field model}

\begin{figure}[tb!]
    \centering
    \includegraphics[width = \linewidth]{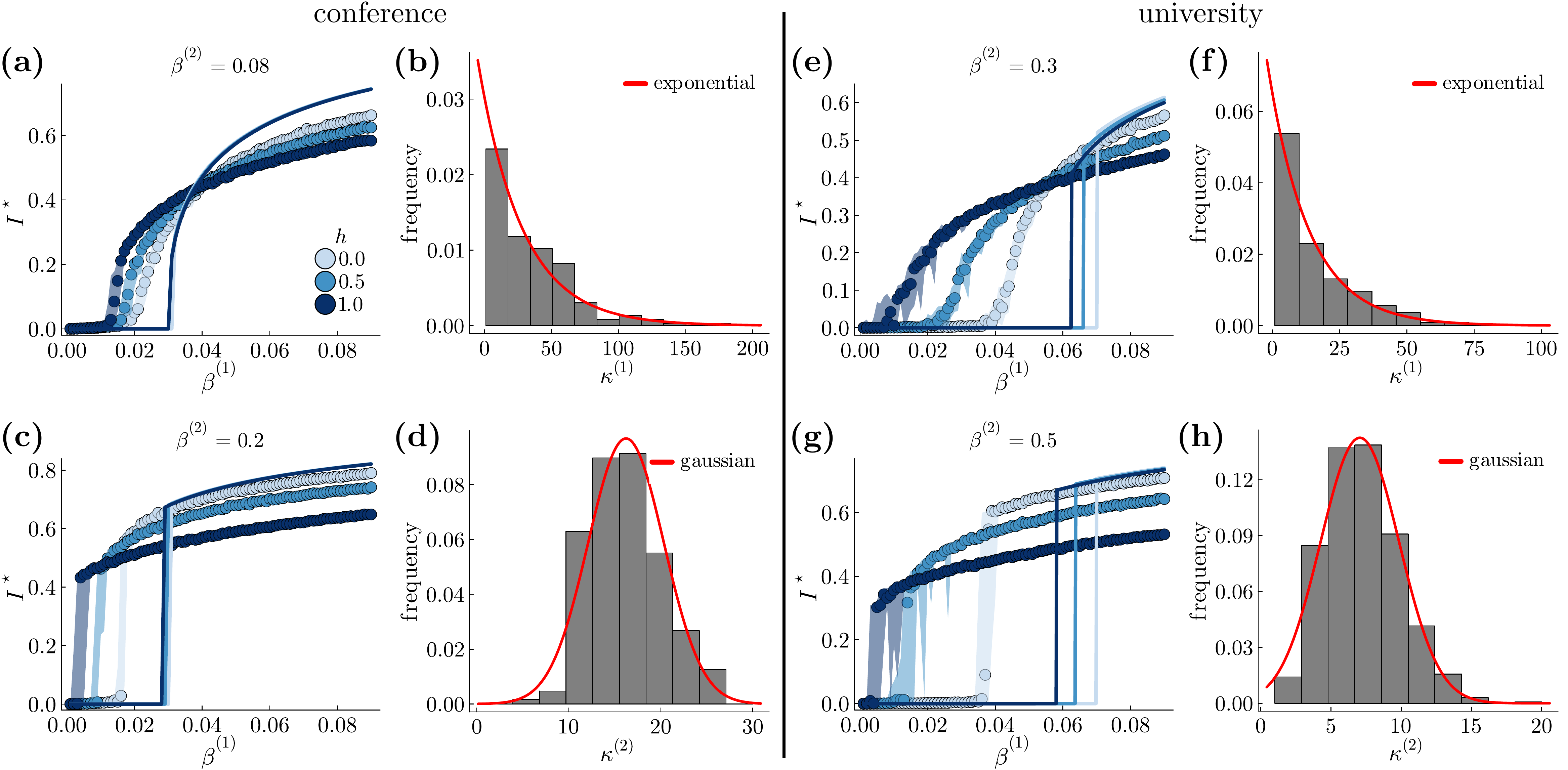}
    \caption{Comparison between the predictions made by the mean-field model (solid lines) and the numerical simulations (points) performed on the hypergraphs constructed from the conference's ((a) and (c)) and the university campus' ((e) and (f)) datasets.
    Points denote medians computed over $20$ random initializations; ribbons cover from the $5$-th to the $95$-th percentile.
    The added three-body interactions form triangles with probability $h=0$ (linear hypergraph), $h=0.5$, and $h=1$ (simplicial complex). ((b) and (d)) $2$-degree and $3$-degree distributions of the hypergraphs generated from the conference's dataset, shown to be well reproduced by an exponential distribution (mean: $1/\langle\kappa^{(1)}\rangle$) and a gaussian distribution (mean: $\langle\kappa^{(2)}\rangle$; variance: $\text{var}(\kappa^{(2)})$), respectively (red curves). ((f) and (h)) As panels (b) and (d), but for the hypergraphs generated from the university campus' dataset.
    }
    \label{fig_1_SM}
\end{figure}

In Fig.~\ref{fig_1_SM} we report the results shown in Fig.~3 of the main text, with the addition of the predictions made by the mean-field model (Eq.~(3)). We also show the $2$-degree and $3$-degree distributions. These are well reproduced by exponential and gaussian distributions, respectively. Given the $2$-degree heterogeneity and assortativity of the generated networks, is no surprise the poor performance of the mean-field approximation, especially in predicting the invasion threshold. The latter is always heavily overestimated. The reasons for this are multiple.

A first reason, valid for any $h$, is the $2$-degree heterogeneity (see panels (b) and (f) in Fig.~\ref{fig_1_SM}). Given that adding three-body interactions can only lower the invasion threshold, an upper bound for the threshold is found considering only two-body interactions, or equivalently, $h=0$ (since, as we proved, $3$-edges not overlapped with $2$-edges do not affect the threshold). A rough estimate of the upper bound is provided by $\langle\kappa^{(1)}\rangle/\langle{\kappa^{(1)}}^2\rangle$, as predicted by a heterogeneous (node-based) mean-field approximation, which yields  a much smaller threshold than a homogeneous mean-field approximation when $\text{var}(\kappa^{(1)})$ is large\footnote{Actually, the prediction made by the heterogeneous mean-field approximation underestimates the threshold. We are indeed dealing with highly clustered networks, for which the threshold is higher than for random networks with the same degree distribution.}.

A second reason is that, on one hand, it has been shown that degree assortativity lowers the threshold~\cite{d2012robustness}; on the other hand, mean-field approximations have been observed to perform exceptionally poorly against assortative networks~\cite{gleeson2012accuracy}. These two facts together strongly suggest an additional contribution to the error made by the mean-field model.

Finally, increasing $h$, this model further overestimates the threshold, for it does not account for triangles sharing two nodes (i.e., a $2$-edge), indeed present in the generated hypergraphs. In fact, once two nodes belonging to $n$ triangles are both infected, $n$ different three-body interactions become simultaneously active. This results in a smaller threshold than the one predicted by the mean-field model, given the sparsity it assumes (i.e., $n=1$ only). To notice that the presence of triangles with shared $2$-edges also explains the much lower prevalence observed for high $h$ compared to that predicted by the mean-field model. Indeed, a set of $n$ triangles sharing a given $2$-edge, involves only $n+2$ nodes. If they shared only one node, the involved nodes would be $2n+1$; if they were all disconnected, the involved nodes would become $3n$. That is, the more frequently triangles share nodes, the more redundant is the structure and, in turn, the smaller are the outbreaks.

It should be noted that a mean-field model is very much expected to fail in providing quantitatively accurate predictions for quenched structures~\cite{kiss2017mathematics}. In fact, the accuracy our mean-field theory shows for the configuration-model hypergraphs in Fig.~2 is not so obvious. For instance, a less refined node-based mean-field model would be very imprecise even in that case, especially in predicting the invasion threshold. In light of this, we can rather appreciate the significant aspect of the results reported in Fig.~\ref{fig_1_SM}: that the phenomenology our theory revealed is, not only still valid for structures violating the assumptions of homogeneity and sparseness, but is actually greatly emphasized!

\subsection{Results for the SIR model}

We show here the results of numerical simulations we performed using a susceptible-infectious-recovered (SIR) contagion model, where upon recovery individuals move to the recovered (R) compartment instead of entering back the susceptible one (S). This higher-order generalization of the SIR model has been very recently analyzed in detail by Lv et al.~\cite{lv2023simplicial}. The authors showed that, as in the SIS model, sufficiently high values of the three-body infection rate, $\beta^{(2)}$, make the phase transition discontinuous in both homogeneous and heterogeneous simplicial complexes. Using however a node-based approximation, their approach is insensible to the way in which two- and three-body interactions are arranged in the structure, hence to their degree of overlap. Consequently, it is not possible to discern a simplicial complex from a linear hypergraph (or any other intermediate structure).

\begin{figure}[t]
    \centering
    \includegraphics[width = 0.85\linewidth]{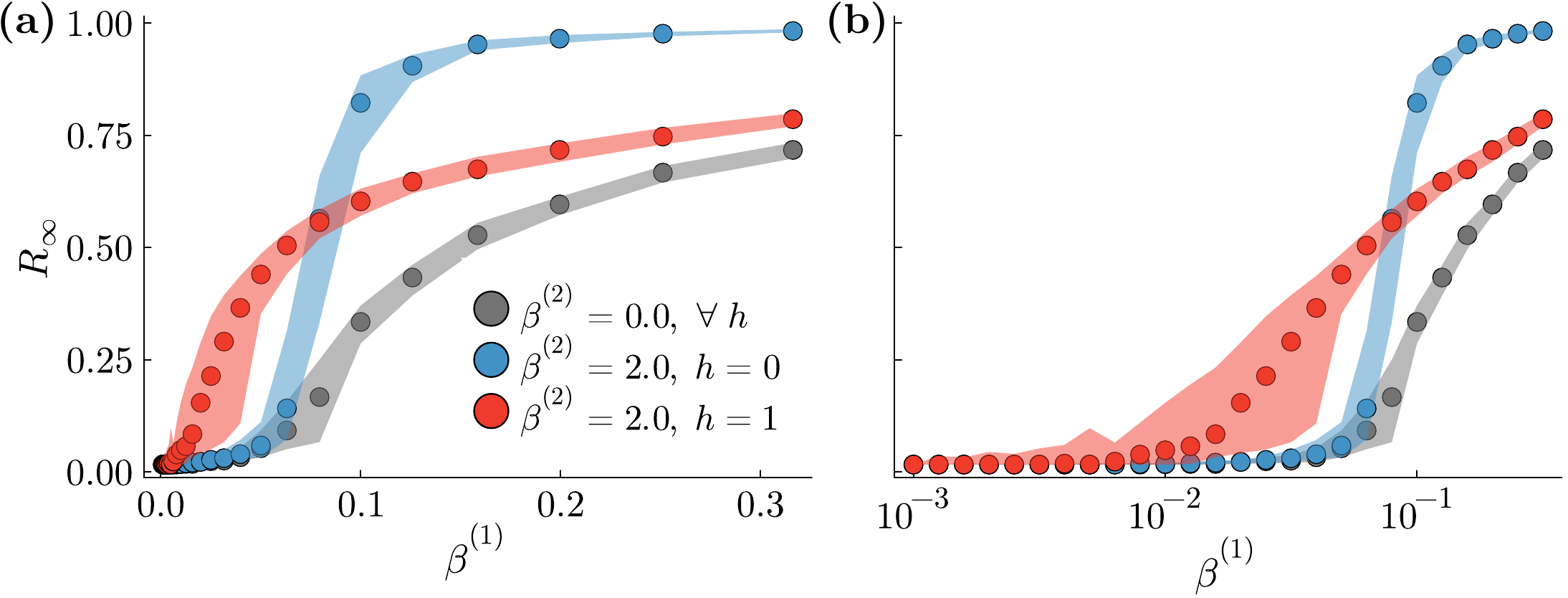}
    \caption{Numerical simulations performed on the hypergraphs constructed from the university campus'~\cite{sapiezynski2019interaction} datasets when the contagion dynamics is a SIR process.
    (a) Final attack rate, $R_\infty$, versus the two-body infection rate, $\beta^{(1)}$. Points denote medians computed over $100$ random initializations; ribbons cover from the $25$-th to the $75$-th percentile.
    The added three-body interactions form triangles with probability $h=0$ (linear hypergraph; blue curve) and $h=1$ (simplicial complex; red curve). Clearly, turning off the three-body interactions ($\beta^{(2)}=0$), varying $h$ has no effect on the dynamics (grey curve).
    (b) Same as panel (a), but with logarithmic abscissa to stretch the low-$\beta^{(1)}$ interval and better appreciate the shift of the invasion threshold.
    }
    \label{fig_2_SM}
\end{figure}

In Fig.~\ref{fig_2_SM}, we show the results for hypergraphs generated from the university campus' dataset. The order parameter for a SIR-like model is the final attack rate, $R_\infty$, which is the total fraction of nodes that got the infection (and eventually recovered) during the entire outbreak (the infected compartment is empty at equilibrium). The simulations confirm the generality of the phenomenology our mean-field model revealed: (i) three-body interactions affect the invasion threshold only if they overlap with two-body interactions ($h>0$); (ii) a larger overlap (higher $h$) implies lower invasion thresholds but also smaller outbreaks; and (iii) varying exclusively the degree of overlap can change the nature of the phase transition. About the last point, it should be noted that the discontinuity of the transition for $\beta^{(2)} = 2$ and $h=0$ is blurred by strong finite-size effects.

\begin{thebibliography}{34}%
\makeatletter
\providecommand \@ifxundefined [1]{%
 \@ifx{#1\undefined}
}%
\providecommand \@ifnum [1]{%
 \ifnum #1\expandafter \@firstoftwo
 \else \expandafter \@secondoftwo
 \fi
}%
\providecommand \@ifx [1]{%
 \ifx #1\expandafter \@firstoftwo
 \else \expandafter \@secondoftwo
 \fi
}%
\providecommand \natexlab [1]{#1}%
\providecommand \enquote  [1]{``#1''}%
\providecommand \bibnamefont  [1]{#1}%
\providecommand \bibfnamefont [1]{#1}%
\providecommand \citenamefont [1]{#1}%
\providecommand \href@noop [0]{\@secondoftwo}%
\providecommand \href [0]{\begingroup \@sanitize@url \@href}%
\providecommand \@href[1]{\@@startlink{#1}\@@href}%
\providecommand \@@href[1]{\endgroup#1\@@endlink}%
\providecommand \@sanitize@url [0]{\catcode `\\12\catcode `\$12\catcode
  `\&12\catcode `\#12\catcode `\^12\catcode `\_12\catcode `\%12\relax}%
\providecommand \@@startlink[1]{}%
\providecommand \@@endlink[0]{}%
\providecommand \url  [0]{\begingroup\@sanitize@url \@url }%
\providecommand \@url [1]{\endgroup\@href {#1}{\urlprefix }}%
\providecommand \urlprefix  [0]{URL }%
\providecommand \Eprint [0]{\href }%
\providecommand \doibase [0]{https://doi.org/}%
\providecommand \selectlanguage [0]{\@gobble}%
\providecommand \bibinfo  [0]{\@secondoftwo}%
\providecommand \bibfield  [0]{\@secondoftwo}%
\providecommand \translation [1]{[#1]}%
\providecommand \BibitemOpen [0]{}%
\providecommand \bibitemStop [0]{}%
\providecommand \bibitemNoStop [0]{.\EOS\space}%
\providecommand \EOS [0]{\spacefactor3000\relax}%
\providecommand \BibitemShut  [1]{\csname bibitem#1\endcsname}%
\let\auto@bib@innerbib\@empty
\bibitem [{\citenamefont {Granovetter}(1978)}]{granovetter1978threshold}%
  \BibitemOpen
  \bibfield  {author} {\bibinfo {author} {\bibfnamefont {M.}~\bibnamefont
  {Granovetter}},\ }\bibfield  {title} {\bibinfo {title} {Threshold models of
  collective behavior},\ }\href@noop {} {\bibfield  {journal} {\bibinfo
  {journal} {American Journal of Sociology}\ }\textbf {\bibinfo {volume}
  {83}},\ \bibinfo {pages} {1420} (\bibinfo {year} {1978})}\BibitemShut
  {NoStop}%
\bibitem [{\citenamefont {Watts}(2002)}]{watts2002simple}%
  \BibitemOpen
  \bibfield  {author} {\bibinfo {author} {\bibfnamefont {D.~J.}\ \bibnamefont
  {Watts}},\ }\bibfield  {title} {\bibinfo {title} {A simple model of global
  cascades on random networks},\ }\href@noop {} {\bibfield  {journal} {\bibinfo
   {journal} {Proceedings of the National Academy of Sciences}\ }\textbf
  {\bibinfo {volume} {99}},\ \bibinfo {pages} {5766} (\bibinfo {year}
  {2002})}\BibitemShut {NoStop}%
\bibitem [{\citenamefont {Dodds}\ and\ \citenamefont
  {Watts}(2004)}]{dodds2004universal}%
  \BibitemOpen
  \bibfield  {author} {\bibinfo {author} {\bibfnamefont {P.~S.}\ \bibnamefont
  {Dodds}}\ and\ \bibinfo {author} {\bibfnamefont {D.~J.}\ \bibnamefont
  {Watts}},\ }\bibfield  {title} {\bibinfo {title} {Universal behavior in a
  generalized model of contagion},\ }\href@noop {} {\bibfield  {journal}
  {\bibinfo  {journal} {Physical Review Letters}\ }\textbf {\bibinfo {volume}
  {92}},\ \bibinfo {pages} {218701} (\bibinfo {year} {2004})}\BibitemShut
  {NoStop}%
\bibitem [{\citenamefont {Centola}\ and\ \citenamefont
  {Macy}(2007)}]{centola2007complex}%
  \BibitemOpen
  \bibfield  {author} {\bibinfo {author} {\bibfnamefont {D.}~\bibnamefont
  {Centola}}\ and\ \bibinfo {author} {\bibfnamefont {M.}~\bibnamefont {Macy}},\
  }\bibfield  {title} {\bibinfo {title} {Complex contagions and the weakness of
  long ties},\ }\href@noop {} {\bibfield  {journal} {\bibinfo  {journal}
  {American Journal of Sociology}\ }\textbf {\bibinfo {volume} {113}},\
  \bibinfo {pages} {702} (\bibinfo {year} {2007})}\BibitemShut {NoStop}%
\bibitem [{\citenamefont {Guilbeault}\ \emph {et~al.}(2018)\citenamefont
  {Guilbeault}, \citenamefont {Becker},\ and\ \citenamefont
  {Centola}}]{Guilbeault2018}%
  \BibitemOpen
  \bibfield  {author} {\bibinfo {author} {\bibfnamefont {D.}~\bibnamefont
  {Guilbeault}}, \bibinfo {author} {\bibfnamefont {J.}~\bibnamefont {Becker}},\
  and\ \bibinfo {author} {\bibfnamefont {D.}~\bibnamefont {Centola}},\
  }\bibinfo {title} {Complex contagions: A decade in review},\ in\ \href@noop
  {} {\emph {\bibinfo {booktitle} {Complex Spreading Phenomena in Social
  Systems: Influence and Contagion in Real-World Social Networks}}}\ (\bibinfo
  {publisher} {Springer International Publishing},\ \bibinfo {address} {Cham},\
  \bibinfo {year} {2018})\ pp.\ \bibinfo {pages} {3--25}\BibitemShut {NoStop}%
\bibitem [{\citenamefont {Romanczuk}\ and\ \citenamefont
  {Daniels}(2023)}]{romanczuk2023phase}%
  \BibitemOpen
  \bibfield  {author} {\bibinfo {author} {\bibfnamefont {P.}~\bibnamefont
  {Romanczuk}}\ and\ \bibinfo {author} {\bibfnamefont {B.~C.}\ \bibnamefont
  {Daniels}},\ }\bibinfo {title} {Phase transitions and criticality in the
  collective behavior of animals—self-organization and biological function},\
  in\ \href@noop {} {\emph {\bibinfo {booktitle} {Order, Disorder and
  Criticality: Advanced Problems of Phase Transition Theory}}}\ (\bibinfo
  {publisher} {World Scientific},\ \bibinfo {year} {2023})\ pp.\ \bibinfo
  {pages} {179--208}\BibitemShut {NoStop}%
\bibitem [{\citenamefont {Goffman}\ and\ \citenamefont
  {Newill}(1964)}]{goffman1964generalization}%
  \BibitemOpen
  \bibfield  {author} {\bibinfo {author} {\bibfnamefont {W.}~\bibnamefont
  {Goffman}}\ and\ \bibinfo {author} {\bibfnamefont {V.}~\bibnamefont
  {Newill}},\ }\bibfield  {title} {\bibinfo {title} {Generalization of epidemic
  theory},\ }\href@noop {} {\bibfield  {journal} {\bibinfo  {journal} {Nature}\
  }\textbf {\bibinfo {volume} {204}},\ \bibinfo {pages} {225} (\bibinfo {year}
  {1964})}\BibitemShut {NoStop}%
\bibitem [{\citenamefont {Daley}\ and\ \citenamefont
  {Kendall}(1964)}]{daley1964epidemics}%
  \BibitemOpen
  \bibfield  {author} {\bibinfo {author} {\bibfnamefont {D.~J.}\ \bibnamefont
  {Daley}}\ and\ \bibinfo {author} {\bibfnamefont {D.~G.}\ \bibnamefont
  {Kendall}},\ }\bibfield  {title} {\bibinfo {title} {Epidemics and rumours},\
  }\href@noop {} {\bibfield  {journal} {\bibinfo  {journal} {Nature}\ }\textbf
  {\bibinfo {volume} {204}},\ \bibinfo {pages} {1118} (\bibinfo {year}
  {1964})}\BibitemShut {NoStop}%
\bibitem [{\citenamefont {Levy}\ and\ \citenamefont
  {Nail}(1993)}]{levy1993contagion}%
  \BibitemOpen
  \bibfield  {author} {\bibinfo {author} {\bibfnamefont {D.~A.}\ \bibnamefont
  {Levy}}\ and\ \bibinfo {author} {\bibfnamefont {P.~R.}\ \bibnamefont
  {Nail}},\ }\bibfield  {title} {\bibinfo {title} {Contagion: a theoretical and
  empirical review and reconceptualization.},\ }\href@noop {} {\bibfield
  {journal} {\bibinfo  {journal} {Genetic, social, and general psychology
  monographs}\ }\textbf {\bibinfo {volume} {119}},\ \bibinfo {pages} {233}
  (\bibinfo {year} {1993})}\BibitemShut {NoStop}%
\bibitem [{\citenamefont {Hill}\ \emph {et~al.}(2010)\citenamefont {Hill},
  \citenamefont {Rand}, \citenamefont {Nowak},\ and\ \citenamefont
  {Christakis}}]{hill2010infectious}%
  \BibitemOpen
  \bibfield  {author} {\bibinfo {author} {\bibfnamefont {A.~L.}\ \bibnamefont
  {Hill}}, \bibinfo {author} {\bibfnamefont {D.~G.}\ \bibnamefont {Rand}},
  \bibinfo {author} {\bibfnamefont {M.~A.}\ \bibnamefont {Nowak}},\ and\
  \bibinfo {author} {\bibfnamefont {N.~A.}\ \bibnamefont {Christakis}},\
  }\bibfield  {title} {\bibinfo {title} {Infectious disease modeling of social
  contagion in networks},\ }\href@noop {} {\bibfield  {journal} {\bibinfo
  {journal} {PLOS computational biology}\ }\textbf {\bibinfo {volume} {6}},\
  \bibinfo {pages} {e1000968} (\bibinfo {year} {2010})}\BibitemShut {NoStop}%
\bibitem [{\citenamefont {Centola}(2010)}]{centola2010spread}%
  \BibitemOpen
  \bibfield  {author} {\bibinfo {author} {\bibfnamefont {D.}~\bibnamefont
  {Centola}},\ }\bibfield  {title} {\bibinfo {title} {The spread of behavior in
  an online social network experiment},\ }\href@noop {} {\bibfield  {journal}
  {\bibinfo  {journal} {Science}\ }\textbf {\bibinfo {volume} {329}},\ \bibinfo
  {pages} {1194} (\bibinfo {year} {2010})}\BibitemShut {NoStop}%
\bibitem [{\citenamefont {O'Sullivan}\ \emph {et~al.}(2015)\citenamefont
  {O'Sullivan}, \citenamefont {O'Keeffe}, \citenamefont {Fennell},\ and\
  \citenamefont {Gleeson}}]{o2015mathematical}%
  \BibitemOpen
  \bibfield  {author} {\bibinfo {author} {\bibfnamefont {D.~J.}\ \bibnamefont
  {O'Sullivan}}, \bibinfo {author} {\bibfnamefont {G.~J.}\ \bibnamefont
  {O'Keeffe}}, \bibinfo {author} {\bibfnamefont {P.~G.}\ \bibnamefont
  {Fennell}},\ and\ \bibinfo {author} {\bibfnamefont {J.~P.}\ \bibnamefont
  {Gleeson}},\ }\bibfield  {title} {\bibinfo {title} {Mathematical modeling of
  complex contagion on clustered networks},\ }\href@noop {} {\bibfield
  {journal} {\bibinfo  {journal} {Frontiers in Physics}\ }\textbf {\bibinfo
  {volume} {3}},\ \bibinfo {pages} {71} (\bibinfo {year} {2015})}\BibitemShut
  {NoStop}%
\bibitem [{\citenamefont {Guilbeault}\ and\ \citenamefont
  {Centola}(2021)}]{guilbeault2021topological}%
  \BibitemOpen
  \bibfield  {author} {\bibinfo {author} {\bibfnamefont {D.}~\bibnamefont
  {Guilbeault}}\ and\ \bibinfo {author} {\bibfnamefont {D.}~\bibnamefont
  {Centola}},\ }\bibfield  {title} {\bibinfo {title} {Topological measures for
  identifying and predicting the spread of complex contagions},\ }\href@noop {}
  {\bibfield  {journal} {\bibinfo  {journal} {Nature communications}\ }\textbf
  {\bibinfo {volume} {12}},\ \bibinfo {pages} {4430} (\bibinfo {year}
  {2021})}\BibitemShut {NoStop}%
\bibitem [{\citenamefont {Iacopini}\ \emph {et~al.}(2019)\citenamefont
  {Iacopini}, \citenamefont {Petri}, \citenamefont {Barrat},\ and\
  \citenamefont {Latora}}]{iacopini2019simplicial}%
  \BibitemOpen
  \bibfield  {author} {\bibinfo {author} {\bibfnamefont {I.}~\bibnamefont
  {Iacopini}}, \bibinfo {author} {\bibfnamefont {G.}~\bibnamefont {Petri}},
  \bibinfo {author} {\bibfnamefont {A.}~\bibnamefont {Barrat}},\ and\ \bibinfo
  {author} {\bibfnamefont {V.}~\bibnamefont {Latora}},\ }\bibfield  {title}
  {\bibinfo {title} {Simplicial models of social contagion},\ }\href@noop {}
  {\bibfield  {journal} {\bibinfo  {journal} {Nature communications}\ }\textbf
  {\bibinfo {volume} {10}},\ \bibinfo {pages} {2485} (\bibinfo {year}
  {2019})}\BibitemShut {NoStop}%
\bibitem [{\citenamefont {Klamt}\ \emph {et~al.}(2009)\citenamefont {Klamt},
  \citenamefont {Haus},\ and\ \citenamefont {Theis}}]{klamt2009hypergraphs}%
  \BibitemOpen
  \bibfield  {author} {\bibinfo {author} {\bibfnamefont {S.}~\bibnamefont
  {Klamt}}, \bibinfo {author} {\bibfnamefont {U.-U.}\ \bibnamefont {Haus}},\
  and\ \bibinfo {author} {\bibfnamefont {F.}~\bibnamefont {Theis}},\ }\bibfield
   {title} {\bibinfo {title} {Hypergraphs and cellular networks},\ }\href@noop
  {} {\bibfield  {journal} {\bibinfo  {journal} {PLoS computational biology}\
  }\textbf {\bibinfo {volume} {5}},\ \bibinfo {pages} {e1000385} (\bibinfo
  {year} {2009})}\BibitemShut {NoStop}%
\bibitem [{\citenamefont {Jost}\ and\ \citenamefont
  {Mulas}(2019)}]{jost2019hypergraph}%
  \BibitemOpen
  \bibfield  {author} {\bibinfo {author} {\bibfnamefont {J.}~\bibnamefont
  {Jost}}\ and\ \bibinfo {author} {\bibfnamefont {R.}~\bibnamefont {Mulas}},\
  }\bibfield  {title} {\bibinfo {title} {Hypergraph laplace operators for
  chemical reaction networks},\ }\href@noop {} {\bibfield  {journal} {\bibinfo
  {journal} {Advances in mathematics}\ }\textbf {\bibinfo {volume} {351}},\
  \bibinfo {pages} {870} (\bibinfo {year} {2019})}\BibitemShut {NoStop}%
\bibitem [{\citenamefont {Tarde}(1890)}]{tarde1890laws}%
  \BibitemOpen
  \bibfield  {author} {\bibinfo {author} {\bibfnamefont {G.}~\bibnamefont
  {Tarde}},\ }\href@noop {} {\emph {\bibinfo {title} {The Laws of Imitation}}}\
  (\bibinfo  {publisher} {Henry Holt and Company},\ \bibinfo {address} {New
  York},\ \bibinfo {year} {1890})\BibitemShut {NoStop}%
\bibitem [{\citenamefont {Blumer}(1939)}]{blumer1939}%
  \BibitemOpen
  \bibfield  {author} {\bibinfo {author} {\bibfnamefont {H.}~\bibnamefont
  {Blumer}},\ }\bibinfo {title} {Collective behavior},\ in\ \href@noop {}
  {\emph {\bibinfo {booktitle} {Principles of Sociology}}},\ \bibinfo {editor}
  {edited by\ \bibinfo {editor} {\bibfnamefont {R.~E.}\ \bibnamefont {Park}}}\
  (\bibinfo  {publisher} {New York, Barnes \& Noble},\ \bibinfo {year} {1939})\
  pp.\ \bibinfo {pages} {219--288}\BibitemShut {NoStop}%
\bibitem [{\citenamefont {Le~Bon}(2002)}]{le2002crowd}%
  \BibitemOpen
  \bibfield  {author} {\bibinfo {author} {\bibfnamefont {G.}~\bibnamefont
  {Le~Bon}},\ }\href@noop {} {\emph {\bibinfo {title} {The crowd: A study of
  the popular mind}}}\ (\bibinfo  {publisher} {Courier Corporation},\ \bibinfo
  {year} {2002})\BibitemShut {NoStop}%
\bibitem [{\citenamefont {Landry}\ and\ \citenamefont
  {Restrepo}(2020)}]{landry2020effect}%
  \BibitemOpen
  \bibfield  {author} {\bibinfo {author} {\bibfnamefont {N.~W.}\ \bibnamefont
  {Landry}}\ and\ \bibinfo {author} {\bibfnamefont {J.~G.}\ \bibnamefont
  {Restrepo}},\ }\bibfield  {title} {\bibinfo {title} {The effect of
  heterogeneity on hypergraph contagion models},\ }\href@noop {} {\bibfield
  {journal} {\bibinfo  {journal} {Chaos: An Interdisciplinary Journal of
  Nonlinear Science}\ }\textbf {\bibinfo {volume} {30}},\ \bibinfo {pages}
  {103117} (\bibinfo {year} {2020})}\BibitemShut {NoStop}%
\bibitem [{\citenamefont {Burgio}\ \emph {et~al.}(2021)\citenamefont {Burgio},
  \citenamefont {Arenas}, \citenamefont {G{\'o}mez},\ and\ \citenamefont
  {Matamalas}}]{burgio2021network}%
  \BibitemOpen
  \bibfield  {author} {\bibinfo {author} {\bibfnamefont {G.}~\bibnamefont
  {Burgio}}, \bibinfo {author} {\bibfnamefont {A.}~\bibnamefont {Arenas}},
  \bibinfo {author} {\bibfnamefont {S.}~\bibnamefont {G{\'o}mez}},\ and\
  \bibinfo {author} {\bibfnamefont {J.~T.}\ \bibnamefont {Matamalas}},\
  }\bibfield  {title} {\bibinfo {title} {Network clique cover approximation to
  analyze complex contagions through group interactions},\ }\href@noop {}
  {\bibfield  {journal} {\bibinfo  {journal} {Communications Physics}\ }\textbf
  {\bibinfo {volume} {4}},\ \bibinfo {pages} {111} (\bibinfo {year}
  {2021})}\BibitemShut {NoStop}%
\bibitem [{\citenamefont {Battiston}\ \emph {et~al.}(2020)\citenamefont
  {Battiston}, \citenamefont {Cencetti}, \citenamefont {Iacopini},
  \citenamefont {Latora}, \citenamefont {Lucas}, \citenamefont {Patania},
  \citenamefont {Young},\ and\ \citenamefont {Petri}}]{battiston2020networks}%
  \BibitemOpen
  \bibfield  {author} {\bibinfo {author} {\bibfnamefont {F.}~\bibnamefont
  {Battiston}}, \bibinfo {author} {\bibfnamefont {G.}~\bibnamefont {Cencetti}},
  \bibinfo {author} {\bibfnamefont {I.}~\bibnamefont {Iacopini}}, \bibinfo
  {author} {\bibfnamefont {V.}~\bibnamefont {Latora}}, \bibinfo {author}
  {\bibfnamefont {M.}~\bibnamefont {Lucas}}, \bibinfo {author} {\bibfnamefont
  {A.}~\bibnamefont {Patania}}, \bibinfo {author} {\bibfnamefont {J.-G.}\
  \bibnamefont {Young}},\ and\ \bibinfo {author} {\bibfnamefont
  {G.}~\bibnamefont {Petri}},\ }\bibfield  {title} {\bibinfo {title} {Networks
  beyond pairwise interactions: structure and dynamics},\ }\href@noop {}
  {\bibfield  {journal} {\bibinfo  {journal} {Physics Reports}\ }\textbf
  {\bibinfo {volume} {874}},\ \bibinfo {pages} {1} (\bibinfo {year}
  {2020})}\BibitemShut {NoStop}%
\bibitem [{\citenamefont {Kiss}\ \emph {et~al.}(2017)\citenamefont {Kiss},
  \citenamefont {Miller}, \citenamefont {Simon} \emph
  {et~al.}}]{kiss2017mathematics}%
  \BibitemOpen
  \bibfield  {author} {\bibinfo {author} {\bibfnamefont {I.~Z.}\ \bibnamefont
  {Kiss}}, \bibinfo {author} {\bibfnamefont {J.~C.}\ \bibnamefont {Miller}},
  \bibinfo {author} {\bibfnamefont {P.~L.}\ \bibnamefont {Simon}}, \emph
  {et~al.},\ }\bibfield  {title} {\bibinfo {title} {Mathematics of epidemics on
  networks},\ }\href@noop {} {\bibfield  {journal} {\bibinfo  {journal} {Cham:
  Springer}\ }\textbf {\bibinfo {volume} {598}},\ \bibinfo {pages} {31}
  (\bibinfo {year} {2017})}\BibitemShut {NoStop}%
\bibitem [{\citenamefont {Bretto}(2013)}]{bretto2013hypergraph}%
  \BibitemOpen
  \bibfield  {author} {\bibinfo {author} {\bibfnamefont {A.}~\bibnamefont
  {Bretto}},\ }\bibfield  {title} {\bibinfo {title} {Hypergraph theory},\
  }\href@noop {} {\bibfield  {journal} {\bibinfo  {journal} {An introduction.
  Mathematical Engineering. Cham: Springer}\ } (\bibinfo {year}
  {2013})}\BibitemShut {NoStop}%
\bibitem [{\citenamefont {Miller}(2009)}]{miller2009percolation}%
  \BibitemOpen
  \bibfield  {author} {\bibinfo {author} {\bibfnamefont {J.~C.}\ \bibnamefont
  {Miller}},\ }\bibfield  {title} {\bibinfo {title} {Percolation and epidemics
  in random clustered networks},\ }\href@noop {} {\bibfield  {journal}
  {\bibinfo  {journal} {Physical Review E}\ }\textbf {\bibinfo {volume} {80}},\
  \bibinfo {pages} {020901} (\bibinfo {year} {2009})}\BibitemShut {NoStop}%
\bibitem [{\citenamefont {H{\'e}bert-Dufresne}\ \emph
  {et~al.}(2010)\citenamefont {H{\'e}bert-Dufresne}, \citenamefont {No{\"e}l},
  \citenamefont {Marceau}, \citenamefont {Allard},\ and\ \citenamefont
  {Dub{\'e}}}]{hebert2010propagation}%
  \BibitemOpen
  \bibfield  {author} {\bibinfo {author} {\bibfnamefont {L.}~\bibnamefont
  {H{\'e}bert-Dufresne}}, \bibinfo {author} {\bibfnamefont {P.-A.}\
  \bibnamefont {No{\"e}l}}, \bibinfo {author} {\bibfnamefont {V.}~\bibnamefont
  {Marceau}}, \bibinfo {author} {\bibfnamefont {A.}~\bibnamefont {Allard}},\
  and\ \bibinfo {author} {\bibfnamefont {L.~J.}\ \bibnamefont {Dub{\'e}}},\
  }\bibfield  {title} {\bibinfo {title} {Propagation dynamics on networks
  featuring complex topologies},\ }\href@noop {} {\bibfield  {journal}
  {\bibinfo  {journal} {Physical Review E}\ }\textbf {\bibinfo {volume} {82}},\
  \bibinfo {pages} {036115} (\bibinfo {year} {2010})}\BibitemShut {NoStop}%
\bibitem [{\citenamefont {Keating}\ \emph {et~al.}(2022)\citenamefont
  {Keating}, \citenamefont {Gleeson},\ and\ \citenamefont
  {O'Sullivan}}]{keating2022multitype}%
  \BibitemOpen
  \bibfield  {author} {\bibinfo {author} {\bibfnamefont {L.~A.}\ \bibnamefont
  {Keating}}, \bibinfo {author} {\bibfnamefont {J.~P.}\ \bibnamefont
  {Gleeson}},\ and\ \bibinfo {author} {\bibfnamefont {D.~J.}\ \bibnamefont
  {O'Sullivan}},\ }\bibfield  {title} {\bibinfo {title} {Multitype branching
  process method for modeling complex contagion on clustered networks},\
  }\href@noop {} {\bibfield  {journal} {\bibinfo  {journal} {Physical Review
  E}\ }\textbf {\bibinfo {volume} {105}},\ \bibinfo {pages} {034306} (\bibinfo
  {year} {2022})}\BibitemShut {NoStop}%
\bibitem [{\citenamefont {G{\'e}nois}\ and\ \citenamefont
  {Barrat}(2018)}]{genois2018can}%
  \BibitemOpen
  \bibfield  {author} {\bibinfo {author} {\bibfnamefont {M.}~\bibnamefont
  {G{\'e}nois}}\ and\ \bibinfo {author} {\bibfnamefont {A.}~\bibnamefont
  {Barrat}},\ }\bibfield  {title} {\bibinfo {title} {Can co-location be used as
  a proxy for face-to-face contacts?},\ }\href@noop {} {\bibfield  {journal}
  {\bibinfo  {journal} {EPJ Data Science}\ }\textbf {\bibinfo {volume} {7}},\
  \bibinfo {pages} {1} (\bibinfo {year} {2018})}\BibitemShut {NoStop}%
\bibitem [{\citenamefont {Sapiezynski}\ \emph {et~al.}(2019)\citenamefont
  {Sapiezynski}, \citenamefont {Stopczynski}, \citenamefont {Lassen},\ and\
  \citenamefont {Lehmann}}]{sapiezynski2019interaction}%
  \BibitemOpen
  \bibfield  {author} {\bibinfo {author} {\bibfnamefont {P.}~\bibnamefont
  {Sapiezynski}}, \bibinfo {author} {\bibfnamefont {A.}~\bibnamefont
  {Stopczynski}}, \bibinfo {author} {\bibfnamefont {D.~D.}\ \bibnamefont
  {Lassen}},\ and\ \bibinfo {author} {\bibfnamefont {S.}~\bibnamefont
  {Lehmann}},\ }\bibfield  {title} {\bibinfo {title} {Interaction data from the
  copenhagen networks study},\ }\href@noop {} {\bibfield  {journal} {\bibinfo
  {journal} {Scientific Data}\ }\textbf {\bibinfo {volume} {6}},\ \bibinfo
  {pages} {315} (\bibinfo {year} {2019})}\BibitemShut {NoStop}%
\bibitem [{\citenamefont {D'Agostino}\ \emph {et~al.}(2012)\citenamefont
  {D'Agostino}, \citenamefont {Scala}, \citenamefont {Zlati{\'c}},\ and\
  \citenamefont {Caldarelli}}]{d2012robustness}%
  \BibitemOpen
  \bibfield  {author} {\bibinfo {author} {\bibfnamefont {G.}~\bibnamefont
  {D'Agostino}}, \bibinfo {author} {\bibfnamefont {A.}~\bibnamefont {Scala}},
  \bibinfo {author} {\bibfnamefont {V.}~\bibnamefont {Zlati{\'c}}},\ and\
  \bibinfo {author} {\bibfnamefont {G.}~\bibnamefont {Caldarelli}},\ }\bibfield
   {title} {\bibinfo {title} {Robustness and assortativity for diffusion-like
  processes in scale-free networks},\ }\href@noop {} {\bibfield  {journal}
  {\bibinfo  {journal} {Europhysics Letters}\ }\textbf {\bibinfo {volume}
  {97}},\ \bibinfo {pages} {68006} (\bibinfo {year} {2012})}\BibitemShut
  {NoStop}%
\bibitem [{\citenamefont {Gleeson}\ \emph {et~al.}(2012)\citenamefont
  {Gleeson}, \citenamefont {Melnik}, \citenamefont {Ward}, \citenamefont
  {Porter},\ and\ \citenamefont {Mucha}}]{gleeson2012accuracy}%
  \BibitemOpen
  \bibfield  {author} {\bibinfo {author} {\bibfnamefont {J.~P.}\ \bibnamefont
  {Gleeson}}, \bibinfo {author} {\bibfnamefont {S.}~\bibnamefont {Melnik}},
  \bibinfo {author} {\bibfnamefont {J.~A.}\ \bibnamefont {Ward}}, \bibinfo
  {author} {\bibfnamefont {M.~A.}\ \bibnamefont {Porter}},\ and\ \bibinfo
  {author} {\bibfnamefont {P.~J.}\ \bibnamefont {Mucha}},\ }\bibfield  {title}
  {\bibinfo {title} {Accuracy of mean-field theory for dynamics on real-world
  networks},\ }\href@noop {} {\bibfield  {journal} {\bibinfo  {journal}
  {Physical Review E}\ }\textbf {\bibinfo {volume} {85}},\ \bibinfo {pages}
  {026106} (\bibinfo {year} {2012})}\BibitemShut {NoStop}%
\bibitem [{\citenamefont {Lv}\ \emph {et~al.}(2023)\citenamefont {Lv},
  \citenamefont {Fan}, \citenamefont {Li}, \citenamefont {Wang},\ and\
  \citenamefont {Zhou}}]{lv2023simplicial}%
  \BibitemOpen
  \bibfield  {author} {\bibinfo {author} {\bibfnamefont {X.}~\bibnamefont
  {Lv}}, \bibinfo {author} {\bibfnamefont {D.}~\bibnamefont {Fan}}, \bibinfo
  {author} {\bibfnamefont {Q.}~\bibnamefont {Li}}, \bibinfo {author}
  {\bibfnamefont {J.}~\bibnamefont {Wang}},\ and\ \bibinfo {author}
  {\bibfnamefont {L.}~\bibnamefont {Zhou}},\ }\bibfield  {title} {\bibinfo
  {title} {Simplicial sir rumor propagation models with delay in both
  homogeneous and heterogeneous networks},\ }\href@noop {} {\bibfield
  {journal} {\bibinfo  {journal} {Physica A: Statistical Mechanics and its
  Applications}\ }\textbf {\bibinfo {volume} {627}},\ \bibinfo {pages} {129131}
  (\bibinfo {year} {2023})}\BibitemShut {NoStop}%
\bibitem [{\citenamefont {Newman}(2002)}]{newman2002assortative}%
  \BibitemOpen
  \bibfield  {author} {\bibinfo {author} {\bibfnamefont {M.}~\bibnamefont
  {Newman}},\ }\bibfield  {title} {\bibinfo {title} {Assortative mixing in
  networks},\ }\href@noop {} {\bibfield  {journal} {\bibinfo  {journal}
  {Physical Review Letters}\ }\textbf {\bibinfo {volume} {89}},\ \bibinfo
  {pages} {208701} (\bibinfo {year} {2002})}\BibitemShut {NoStop}%
\bibitem [{\citenamefont {Zhang}\ \emph {et~al.}(2023)\citenamefont {Zhang},
  \citenamefont {Lucas},\ and\ \citenamefont {Battiston}}]{zhang2023higher}%
  \BibitemOpen
  \bibfield  {author} {\bibinfo {author} {\bibfnamefont {Y.}~\bibnamefont
  {Zhang}}, \bibinfo {author} {\bibfnamefont {M.}~\bibnamefont {Lucas}},\ and\
  \bibinfo {author} {\bibfnamefont {F.}~\bibnamefont {Battiston}},\ }\bibfield
  {title} {\bibinfo {title} {Higher-order interactions shape collective
  dynamics differently in hypergraphs and simplicial complexes},\ }\href@noop
  {} {\bibfield  {journal} {\bibinfo  {journal} {Nature Communications}\
  }\textbf {\bibinfo {volume} {14}},\ \bibinfo {pages} {1605} (\bibinfo {year}
  {2023})}\BibitemShut {NoStop}%
\end{thebibliography}

%

\end{document}